\documentclass[%
reprint,%
 amssymb, amsmath,%
 aip,apl,%
frontmatterverbose,
]{revtex4-1}	
\usepackage{braket}
\usepackage{bm}%
\usepackage[colorlinks=true,linkcolor=blue]{hyperref}%
\expandafter\ifx\csname package@font\endcsname\relax\else
 \expandafter\expandafter
 \expandafter\usepackage
 \expandafter\expandafter
 \expandafter{\csname package@font\endcsname}%
\fi
\usepackage{graphicx}
\usepackage[utf8]{inputenc}

\begin{document}

\title{Loop-gap Microwave Resonator for Hybrid Quantum Systems}%

\author{Jason R. Ball}%
\affiliation{Okinawa Institute of Science and Technology Graduate University, Onna, Okinawa 904-0495, Japan}%

\author{Yu Yamashiro}%
\affiliation{Okinawa Institute of Science and Technology Graduate University, Onna, Okinawa 904-0495, Japan}%
\altaffiliation[Present address: ]{Department of Physics, Tokyo Institute of Technology, Oh-okayama, Meguro-ku, Tokyo 152-8551, Japan}
\affiliation{Department of Physics, University of the Ryukyus, Nishihara, Okinawa 903-0213, Japan}%

\author{Hitoshi Sumiya}
\affiliation{Advanced Materials Laboratory, Sumitomo Electric Industries Ltd., Itami, Osaka 664-001, Japan}

\author{Shinobu Onoda}%
\affiliation{Takasaki Advanced Radiation Research Institute, National Institutes for Quantum and Radiological Science and Technology, Takasaki, Gunma 370-1292, Japan}%

\author{Takeshi Ohshima}%
\affiliation{Takasaki Advanced Radiation Research Institute, National Institutes for Quantum and Radiological Science and Technology, Takasaki, Gunma 370-1292, Japan}%

\author{Junichi Isoya}
\affiliation{Research Center for Knowledge Communities, University of Tsukuba, Tsukuba, Ibaraki 305-8550, Japan}

\author{Denis Konstantinov}%
\affiliation{Okinawa Institute of Science and Technology Graduate University, Onna, Okinawa 904-0495, Japan}%

\author{Yuimaru Kubo}%
\email{yuimaru.kubo@oist.jp}
\affiliation{Okinawa Institute of Science and Technology Graduate University, Onna, Okinawa 904-0495, Japan}%
\affiliation{PRESTO, Japan Science and Technology (JST), Kawaguchi, Saitama 332-0012, Japan}

\date{\today}%

\begin{abstract}

We designed a loop-gap microwave resonator for applications of spin-based hybrid quantum systems, and tested it with impurity spins in diamond. 
Strong coupling with ensembles of nitrogen-vacancy (NV) centers and substitutional nitrogen (P1) centers was observed. 
These results show that loop-gap resonators are viable in the prospect of spin-based hybrid quantum systems, especially for an ensemble quantum memory or a quantum transducer.  

\end{abstract}

\maketitle



Superconducting quantum circuits have been remarkably developed in the past decade, and are promising candidates for a quantum computer\cite{kelly_state_2015}. 
However, their decoherence times are relatively shorter than those of microscopic systems. 
Moreover, the low energy of microwave photons fundamentally prevents one from transferring quantum information processed in one superconducting quantum computer inside a dilution refrigerator to somewhere else outside of the millikelvin environment. 

Besides superconducting circuits, spins in solid crystals are another promising quantum system at microwave frequencies mainly because of their long decoherence times\cite{tyryshkin_electron_2011, wolfowicz_atomic_2013, balasubramanian_ultralong_2009, mizuochi_coherence_2009, maurer_room-temperature_2012, bar-gill_solid-state_2013}. 
They are considered to be one of the ideal systems to build hybrid quantum systems\cite{wesenberg_quantum_2009, xiang_hybrid_2013, kurizki_quantum_2015}, where one can exploit both the good coherence of the spins and the designability and controllability of superconducting quantum circuits. 
It has been demonstrated that impurity spins in solid crystals are compatible with superconducting circuits: strong coupling with nitrogen-vacancy (NV) centers in diamond\cite{kubo_strong_2010, kubo_hybrid_2011, zhu_coherent_2011, kubo_storage_2012, amsuss_cavity_2011, putz_protecting_2014}, nitrogen (P1) centers in diamond \cite{schuster_high-cooperativity_2010, ranjan_probing_2013}, rare-earth ions in optical crystals\cite{bushev_ultralow-power_2011, probst_anisotropic_2013}, donors in silicon\cite{zollitsch_high_2015}, and magnons\cite{huebl_high_2013, tabuchi_coherent_2015} have been demonstrated. 
The key to fully exploiting the long decoherence times of spins is the spin echo (refocusing) protocol, in which spins have to be inverted by a microwave pulse within a time scale of free induction decay\cite{wu_storage_2010, julsgaard_quantum_2013, grezes_multimode_2014}. 
However, in most previous works spins have not been uniformly driven because of the spacial inhomogeneity of the microwave magnetic fields in two dimensional resonator geometries\cite{probst_microwave_2015, grezes_multimode_2014, sigillito_fast_2014}
This has made it problematic to invert all the spins simultaneously. 
To circumvent this issue, some works using three dimensional microwave resonators have been reported\cite{abe_electron_2011, probst_three-dimensional_2014, abdurakhimov_normal-mode_2015, creedon_strong_2015, angerer_collective_2016, le_floch_towards_2016, rose_coherent_2017}. 

Apart from the microwave aspect, many spin systems also possess optical transitions, which may be exploited to realize a quantum transducer, a device that coherently and bidirectionally converts between microwave and optical photons. 
Indeed, there have been several proposals to realize such a transducer using spins \cite{williamson_magneto-optic_2014, obrien_interfacing_2014, hisatomi_bidirectional_2016, blum_interfacing_2015}. 
One of the crucial requirements of such a quantum transducer is an efficient mode-matching between microwave and optical resonators \cite{williamson_magneto-optic_2014}. 
Here again, two dimensional superconducting resonators are not ideal to this end for two reasons: 
the inhomogeneous microwave magnetic field, and the superconducting gaps which have much less energy than optical photons.

In this Letter, we design and test a loop-gap microwave resonator\cite{hyde_multipurpose_1989, chen_coupling_2016}, and show its viability for spin-based hybrid quantum systems. 
The microwave AC magnetic field is homogeneous ($\approx 93\, \%$) over the sample volume, and most of the magnetic energy ($\approx 94\, \%$) is confined in the sample space. 
While the former would facilitate uniform driving of the spins, 
the latter would enable efficient overlap between the microwave and optical modes if we implement an optical resonator whose mode passes the sample space of the loop-gap resonator \cite{williamson_magneto-optic_2014}.  
Therefore, the loop-gap-resonator is an ideal component for the realization of a quantum transducer, as discussed above and in Ref \cite{williamson_magneto-optic_2014}. 
We demonstrate strong coupling of microwave photons to an ensemble of nitrogen-vacancy (NV) centers and substitutional nitrogen (P1) centers in diamond. 
We also show that the external quality factor can be tuned from $\sim 10^{3}$ to $\sim 10^{5} $. 


\begin{figure}[ht]
\centering
\includegraphics[width=\hsize]{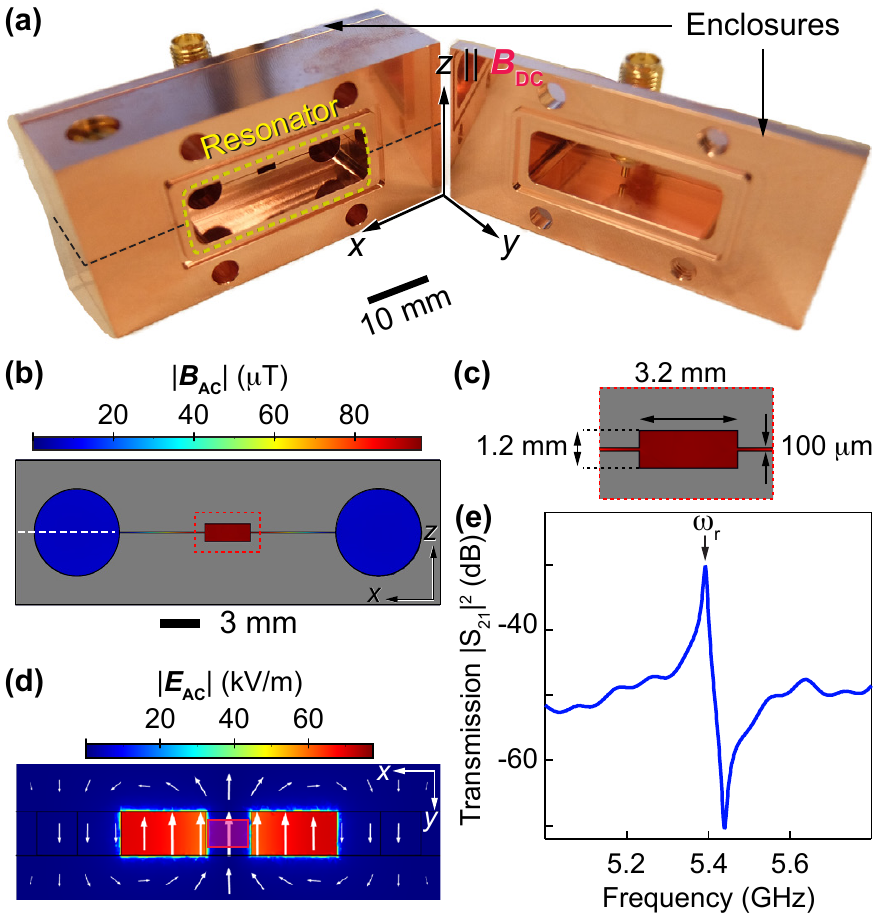}
\caption{The loop-gap microwave resonator designed and tested in this work. 
(a) Photograph of the loop-gap resonator and enclosures. 
Both the resonator and enclosures are made of oxygen free copper.  
An SMA coaxial antenna pin is screwed into each enclosure. 
The black dotted lines are the seam between the two halves which form the resonator. 
(b) A side view of the resonator [$xz$-plane and inside the dashed box in (a)] designed in an electromagnetic simulator. 
Color shows the norm of the AC magnetic field $|\mathbf{B_{\mathrm{AC}}}|$ for an incident microwave power of $- 106\, \mathrm{dBm}$. 
The sample is placed at the center where $|\mathbf{B_{\mathrm{AC}}}|$ is maximum. 
(c) Zoom of the sample space and the gaps [dashed box in (b)].  
(d) A top view [$xy$-plane in (a) at the height of white horizontal dashed line in (b)]. 
Color shows the norm of the AC electric field $|\mathbf{E_{\mathrm{AC}}}|$ in the same condition as in (b), and the light blue arrows are vector representations of the magnetic field mode. 
The length of the arrows corresponds to the strength of the magnetic field (in $\log$ scale). 
(e) An example of transmission spectrum $|S_{21}|$ of the loop-gap resonator measured at room temperature. 
}
\label{fig:1}
\end{figure}

Figure \ref{fig:1} (a) is a photograph of the loop-gap resonator used in this work.  
The resonator is made of oxygen free copper, and consists of a sample space at the center with two $100\,    \mu\mathrm{m}$ gap slits on both sides [Fig.\ref{fig:1}(b) and (c)], forming a lumped-element inductance $L$ and capacitor $C$ respectively, i.e., a three-dimensional \textit{LC} resonator \cite{williamson_magneto-optic_2014, hyde_multipurpose_1989}. 
As shown in Fig. \ref{fig:1}b and c, most of the magnetic field energy ($\approx 94\, \%$ ) of the resonant microwave mode is concentrated in the sample space, yielding the r.m.s. magnetic vacuum field of about $14\, \mathrm{pT}$. 
As a result, the coupling constant $g$ of the microwave photon in the resonator to a single NV center or to a P1 center is about $0.2\, \mathrm{Hz}$ (Supplementary Material), which are lower than the typical value of two dimensional coplanar resonators ($\sim \mathrm{Hz}$\cite{kubo_strong_2010, schuster_high-cooperativity_2010}) but one or two orders of magnitude higher than that of conventional three dimensional resonators ($\sim \mathrm{mHz} - 10\,    \mathrm{mHz}$ \cite{abe_electron_2011, rose_coherent_2017}). 
The resonator was designed to have a resonant frequency of the main loop-gap mode [Fig. \ref{fig:1} (d)] of around $5 - 6\,    \mathrm{GHz}$, such that it would match up with the typical frequency of microwave photons in superconducting quantum circuits. 
A transmission spectrum measured at room temperature is shown in Fig. \ref{fig:1} (e). 
The resonance frequency $\omega_{\mathrm{r}}/2\pi$ is measured to be $5.39\,    \mathrm{GHz}$. 
The asymmetric resonance shape is attributed to the geometry of the assembled resonator and enclosure (Supplementary Material). 

The resonator is capacitively coupled to the input and output antenna pins, each of which are screwed into a copper enclosure (Fig. \ref{fig:1}a). 
As shown later, the external quality factor $Q_{\mathrm{ext}}$ can be tuned by adjusting the depth of the antenna pins. 
The resonator is placed at the mixing chamber in a dilution refrigerator with a base temperature of about $10\,    \mathrm{mK}$, where both the microwave resonator and spins are fully polarized to their lowest energy state. 
An external constant magnetic field ($\mathbf{B_{\mathrm{DC}}}$) is generated by a homemade uniaxial superconducting coil that is thermalized at the still plate of the dilution refrigerator. 
$\mathbf{B_{\mathrm{DC}}}$ is perpendicular to the microwave magnetic field $\mathbf{B_{\mathrm{AC}}}$ [Fig. \ref{fig:1}(a)]. 
The transmission spectra $|S_{21} (\omega)|$ were taken with a vector network analyzer (VNA). 
See supplementary material for the details of the experimental setup.

The two samples investigated in this work are artificial diamond crystals synthesized by HPHT (high-pressure-high-temperature) method. 
Negatively charged NV centers and substitutional nitrogen (P1) centers are contained in the amounts summarized in Table. \ref{table:samples}. 

\begin{table}
\centering
\begin{tabular}{ c c  c  c }
\hline \hline
Sample & Size ($\mathrm{mm}^3$) & [NV] & [P1]  \\
\hline
$\#1$ & $3 \times 1.5 \times 1.1$ & $10\,    \mathrm{ppm}$ & $20\,    \mathrm{ppm}$ \\
\hline
$\#2$ & $3 \times 1.5 \times 0.5$ & $2\,    \mathrm{ppm}$ & $10\,    \mathrm{ppm}$ \\
\hline \hline
\end{tabular}
\caption{Single crystal diamond samples studied in this work. }
\label{table:samples}
\end{table}


The NV center spin Hamiltonian at magnetic fields $\gtrsim 10\,    \mathrm{mT}$ can be written as \cite{felton_hyperfine_2009, acosta_diamonds_2009}
\begin{equation}
H_{NV}/h = \gamma_e \mathbf{B_{\mathrm{DC}}} \mathbf{S} + DS_{z}^2 + \mathbf{S} \mathcal{A}_{\mathrm{N}} \mathbf{I} + \mathcal{P} I_{z}^2,
\end{equation}
where $\gamma_e (= 28\,    \mathrm{MHz/mT})$ is the gyromagnetic ratio of the NV electron spin, $\mathbf{S}$ $(\mathbf{I})$ is the spin operator of the NV electron ($^{14}\mathrm{N}$ nuclear) spin $S = 1$ ($I = 1$), $D = 2.8775\,    \mathrm{GHz}$ is the zero-field splitting between $m_S = 0$ and $m_S = \pm 1$. 
$\mathcal{A}_{\mathrm{N}} (\mathcal{A}_{\mathrm{N}\perp} = -2.7\,     \mathrm{MHz}, \,      \mathcal{A}_{\mathrm{N}\parallel} = -2.1\,     \mathrm{MHz} )$ 
and $\mathcal{P} = -5 \,\mathrm{MHz}$ are the hyperfine interaction tensor and the nuclear quadrupole moment with the $^{14}\mathrm{N}$ nuclear spin, respectively. 
Even though the transverse spin operators $S_x$ and $S_y$ are coupled to the local electric field and strain, which significantly modify the spin eigen states $m_S = \pm 1$ at low magnetic fields \cite{dolde_electric-field_2011, jamonneau_competition_2016}, this effect is negligible at higher magnetic fields where all the measurements were performed in this work. 

\begin{figure}[th]
\centering
\includegraphics[width=\hsize]{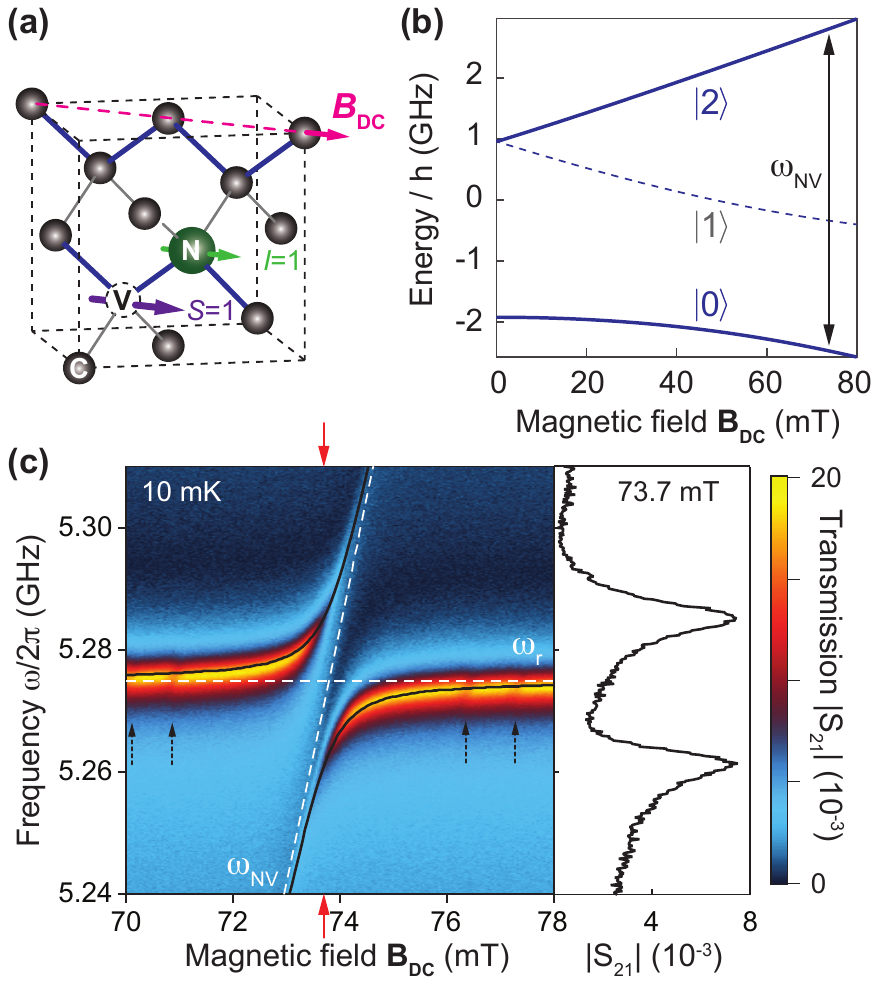}
\caption{NV center results. 
(a) Schematic of an NV center in a diamond crystal with the orientation of the external constant magnetic field $|\mathbf{B_{\mathrm{DC}}}| \parallel [1 1 0]$.  
In this configuration, there are two possible orientations of NV center with respect to $\mathbf{B_{\mathrm{DC}}}$: orthogonal ones (bonds drawn in gray) and non-orthogonal ones (thick blue). 
(b) Energy levels of the non-orthogonal NV centers. 
The transition investigated in this work is between the lowest $|0\rangle$ and the higest $|2\rangle$ states, which are drawn in solid blue curves. 
(c) Transmission spectra color plot of Sample $\#1$ (left), and the spectrum at $73.7\,    \mathrm{mT}$ (right). 
The white dashed lines are the original eigenfrequencies of the NV centers ($\omega_{\mathrm{NV}}$) and the loop-gap resonator ($\omega_{\mathrm{r}}$). 
The dotted arrows on the left panel highlight the other spin transitions due to the hyperfine coupling to $^{13}\mathrm{C}$ nuclei. 
}
\label{fig:2}
\end{figure}

Figure \ref{fig:2} (a) depicts a schematic of an NV center in the diamond lattice and the direction of $\mathbf{B_{\mathrm{DC}}}$. 
Here the case for Sample $\#1$ ($\mathbf{B_{\mathrm{DC}}}\,     \parallel [1 1 0]$) is drawn. 
In this magnetic field orientation, the system can be regarded as two sets of sub-spin ensembles, each of which has a different $\mathbf{B_{\mathrm{DC}}}$ projection on the four possible N-V bond directions: orthogonal and non-orthogonal ones (drawn in thin gray and thick blue respectively). 
Here we focus on the latter one, and its energy level diagram is shown in Fig. \ref{fig:2} (b). 
The magnetic field where the electron spin resonance (ESR) frequency $\omega_{\mathrm{NV}}$ between states $|0\rangle \leftrightarrow |2\rangle$ and the loop-gap resonator $\omega_{\mathrm{r}}$ match is about $74\,    \mathrm{mT}$ (vertical arrow). 

On the left panel in Fig. \ref{fig:2} (c), a color map of transmission spectra $|S_{21} (\omega)|$ versus $\mathbf{B_{\mathrm{DC}}}$ is shown. 
When the ESR frequency $\omega_{\mathrm{NV}}$ (diagonal dashed line) crosses the resonator frequency $\omega_{\mathrm{r}}$ (horizontal dashed line), the spectrum reveals an avoided level crossing, which is the manifestation of the strong coupling between the NV center ensemble and microwave photons in the loop-gap resonator. 
Two polaritonic peaks are observed at $73.7\,    \mathrm{mT}$ (red arrow), and are plotted on the right panel in Fig. \ref{fig:2} (c). 
We fitted the data by a spin ensemble-resonator coupled Hamiltonian\cite{kubo_strong_2010} and calculated the new coupled eigenstates (supplementary material), which are plotted as solid curves. 
The obtained ensemble coupling constant is about $11.5\,    \mathrm{MHz}$, which is in a good agreement with the estimated value (supplementary material). 
In Fig. \ref{fig:2} (c), one can also see absorptions (dotted arrows) away from the avoided level crossing. 
These are ESR transitions of NV centers hyperfine-coupled with the nuclear spins of $^{13}\mathrm{C} \,     \,      (I = 1/2)$ (supplementary material). 

\begin{figure}[t]
\centering
\includegraphics[width=0.9\hsize]{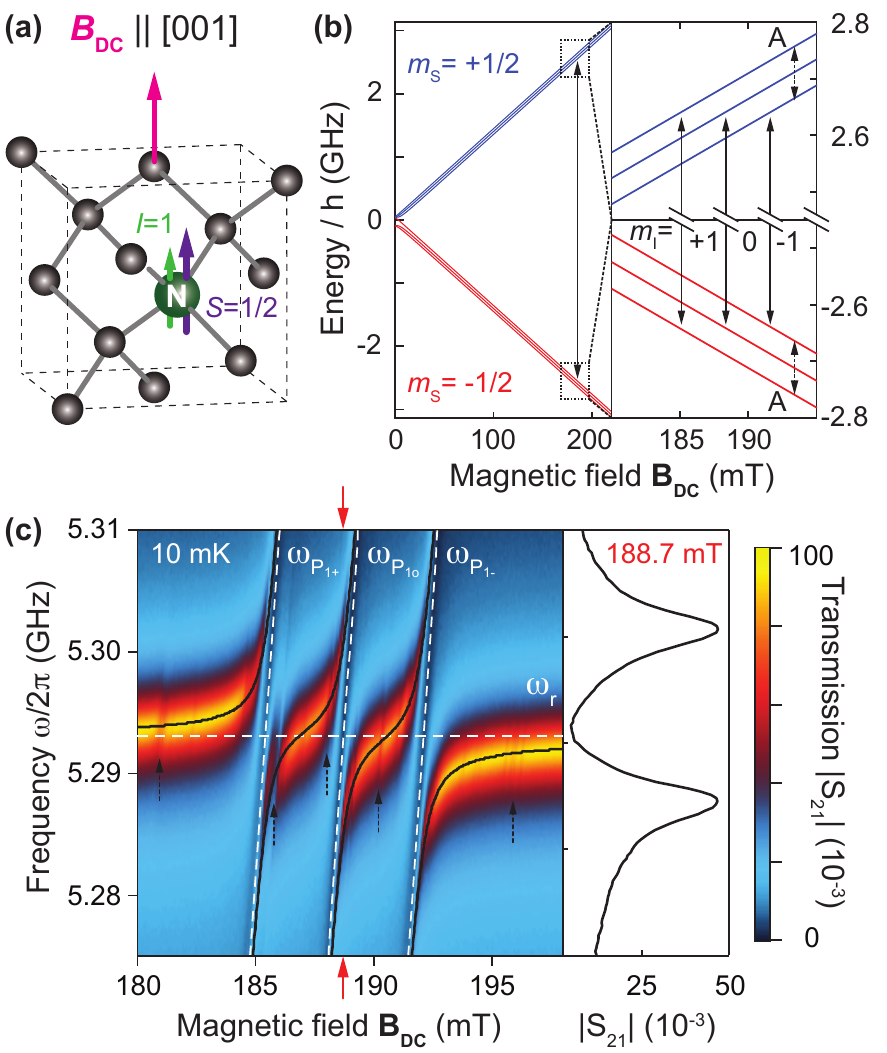}
\caption{P1 center results. 
(a) Schematic of a P1 center in diamond crystal with the orientation of the external constant magnetic field $B_{\mathrm{DC}} \parallel [0 0 1]$ (for Sample $\#2$).  
(b) Energy levels of a P1 center versus $\mathbf{B_{\mathrm{DC}}}$ in this orientation: global (left) and zooms (right). 
The three possible nuclear spin conserving ESR transitions are depicted by the vertical arrows. 
(c) Resonator transmission spectra color plot for Sample $\#2$ (left) and the spectrum at $188.7\,    \mathrm{mT}$ (right).  
The bare frequencies of the ESR transitions for each nuclear spin state ($\omega_{\mathrm{P_{1+}}}$, $\omega_{\mathrm{P_{1o}}}$, and $\omega_{\mathrm{P_{1-}}}$) and the resonator ($\omega_{\mathrm{r}}$) are drawn in white dashed lines.  
The dotted arrows show other ESR transitions of NV centers, or of P1 (NV) centers hyperfine-coupled to $^{13}\mathrm{C}$ nuclei (supplementary material). 
}
\label{fig:3}
\end{figure}


We now focus on P1 (substitutional nitrogen) center, which is another impurity spin in diamond. 
The structure of a P1 center is depicted in Fig. \ref{fig:3} (a). 
It consists of an electron spin $S = 1/2$ and a nuclear spin $I = 1$ of $^{14}$N. 
For the experimental results presented below, we used Sample $\#2$, which has a different crystallographic orientation from Sample $\#1$. 
$\mathbf{B_{\mathrm{DC}}}$ is aligned parallel to $[001]$ as shown in Fig. \ref{fig:3} (a). 
The spin Hamiltonian of a P1 center is written as 
\begin{equation}
H_{P1}/h = \gamma_e \mathbf{B_{\mathrm{DC}}}\mathbf{S}  + \mathbf{S} \mathcal{A}_{\mathrm{N'}} \mathbf{I}, 
\end{equation}
where $\mathbf{S}$ $(\mathbf{I})$ is the spin operator of the P1 electron ($^{14}\mathrm{N}$ nuclear) spin $S = 1/2$ ($I = 1$), and $\mathcal{A}_{\mathrm{N'}}\,      (\mathcal{A}_{\mathrm{N'}\perp} = 114.03\,    \mathrm{MHz}, \mathcal{A}_{\mathrm{N'}\parallel} = 81.33\,    \mathrm{MHz})$ is the hyperfine tensor with the $^{14}\mathrm{N}$ nuclear spin. 
The energy levels of a P1 center versus $\mathbf{B_{\mathrm{DC}}}$ in this orientation are shown in Fig. \ref{fig:3} (b). 
Because of the hyperfine coupling, three possible nuclear spin conserving ESR transitions are allowed in the transition $|m_S = +\frac{1}{2} \rangle \leftrightarrow |- \frac{1}{2} \rangle $, as shown in Fig. \ref{fig:3}(b) by vertical arrows. 

Transmission spectra of the loop-gap resonator with Sample $\# 2$ versus $\mathbf{B_{\mathrm{DC}}}$ is color-plotted in Fig. \ref{fig:3} (c). 
When each ESR frequency (diagonal dashed lines) crosses the resonator frequency $\omega_{\mathrm{r}}$ (horizontal dashed line), an avoided level crossing emerges. 
The ensemble coupling constants are estimated to be $g_{ens,+} = 8.8 \,     \mathrm{MHz}, g_{ens,0} = 7.9 \,     \mathrm{MHz}$, and $g_{ens,-} = 7.8 \,     \mathrm{MHz}$ (supplementary material), where the symbol corresponds to the nuclear spin state $| m_I\rangle$. 
One also finds several other resonances, marked by the dotted arrows. 
Most of them are $^{13}\mathrm{C}$ hyperfine-coupled P1 centers. 
In addition, there are also ESR transitions of NV centers in this magnetic field range (supplementary material).

\begin{figure}[t]
\centering
\includegraphics[width=0.9\hsize]{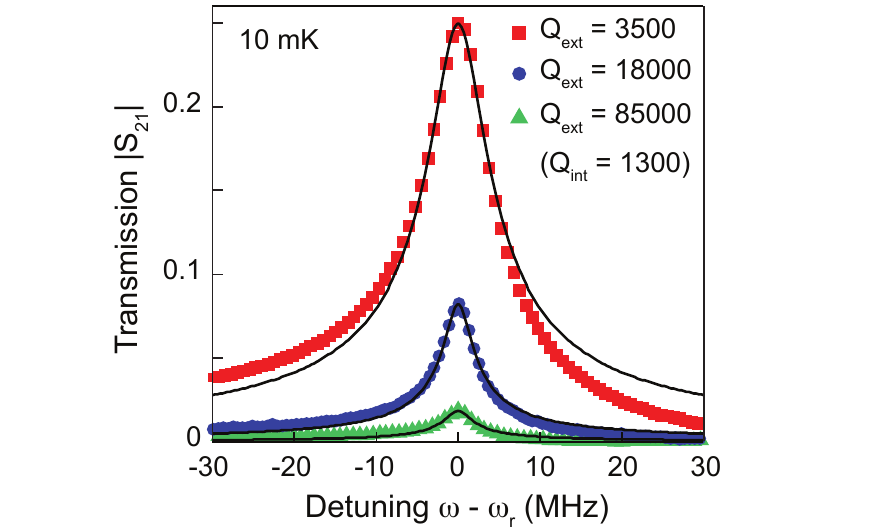}
\caption{External coupling tunability. 
Resonator transmission spectra with different antenna depth are plotted. 
The minimum external quality factor $Q_{\mathrm{ext}}$ is $3500$, while the maximum is $85000$ in the enclosure design used in this work. 
Curves are Lorentzian fit, and symbols are experimental data taken at $10\,    \mathrm{mK}$. 
}
\label{fig:4}
\end{figure}

The ability to tune the external quality factor $Q_{\mathrm{ext}}$ allows the loop-gap resonator to be used for a variety of hybrid quantum systems applications. 
For example, high quality factor is required for quantum information transfer\cite{kubo_hybrid_2011, kubo_storage_2012}, whereas low quality factor is preferred to apply broadband and high power pulses for spin echoes\cite{julsgaard_quantum_2013, grezes_multimode_2014}. 
Fig. \ref{fig:4} presents the tunability of the loop-gap resonator, where we tuned $Q_{\mathrm{ext}}$ from $3500$ to $85000$ by changing the depth of the antennae. 
Pulling them further away would have made the maximum $Q_{\mathrm{ext}}$ even much higher, whereas approaching the pins closer to the loop-gap resonator would make the minimum $Q_{\mathrm{ext}}$ smaller. 
On the other hand, the internal quality factor $Q_{\mathrm{int}}$ is limited to about $1300$. 
We attribute this mainly to the loss from the seam\cite{brecht_demonstration_2015}; the loop-gap resonator is assembled from two symmetric halves [Fig. \ref{fig:1}(a)] without any indium seals. 
Designing the loop-gap resonator in a single piece would avoid the seam loss. 
Another possible loss comes from the surface; we did not perform any electro-mechanical or electro-chemical polish, which may also improve $Q_{\mathrm{int}}$ \cite{angerer_collective_2016}. 

In conclusion, we designed and tested a loop-gap microwave resonator, and demonstrated strong coupling with an NV center and P1 center ensembles in diamond. 
These results show that loop-gap resonators are viable in the prospect of spin-based hybrid quantum systems, especially for a quantum memory or quantum transducer.  

See supplementary material for the design of the resonator and the AC field homogeneity, analysis on the asymmetric resonance using a lumped-element circuit model, calculations of the single-spin and ensemble coupling constants, other ESR transitions in Figs. \ref{fig:2}(c) and \ref{fig:3}(c), the calculation of the new coupled eigenstates, and detailed information about the diamond samples and experimental setup. 

The authors thank S. Ikemiyagi, Y. Ohba, and Y. Higashi in Mechanical Engineering and Microfabrication Support Section at OIST for their technical help, A. Gourmelon for the help on the setup, and D. Vion for the measurement program. 
We acknowledge Y. Tabuchi, T. Shintake, and J. Longdell for their comments on the design of the loop-gap resonators, and P. Bertet and D. Esteve for their comments on this manuscript. 
This work was supported by JST-PRESTO (Grant No. JPMJPR15P7) and OIST Graduate University.

\bibliography{LoopGapResoPaper2017}

\end{document}